\documentclass[sigconf]{acmart}
\usepackage{array}

\usepackage{balance}
\usepackage{booktabs} 
\usepackage{subfigure}

\copyrightyear{2022}
\acmYear{2022}
\setcopyright{acmcopyright}\acmConference[SIGSIM-PADS '22]{SIGSIM Conference on Principles of Advanced Discrete Simulation}{June 8--10, 2022}{Atlanta, GA, USA}
\acmBooktitle{SIGSIM Conference on Principles of Advanced Discrete Simulation (SIGSIM-PADS '22), June 8--10, 2022, Atlanta, GA, USA}
\acmPrice{15.00}
\acmDOI{10.1145/3518997.3534959}
\acmISBN{978-1-4503-9261-7/22/06}

\begin{document}
\title{Resource-Mediated Consensus Formation}

\author{Omar Malik}
\affiliation{%
  \institution{Department of Physics, Applied Physics and Astronomy and the Network Science and Technology Center, Rensselaer Polytechnic Institute}
  \city{Troy}
  \state{NY}
  \country{USA}
  \postcode{12180}
}
\email{maliko@rpi.edu}

\author{James Flamino}
\affiliation{%
  \institution{Network Science and Technology Center, Rensselaer Polytechnic Institute}
  \city{Troy}
  \state{NY}
  \country{USA}
  \postcode{12180}
}
\email{flamij2@rpi.edu}

\author{Boleslaw K. Szymanski}
\affiliation{%
  \institution{Department of Computer Science and the Network Science and Technology Center, Rensselaer Polytechnic Institute}
  \city{Troy}
  \state{NY}
  \country{USA}
  \postcode{12180}
}
\email{szymab@rpi.edu}

\renewcommand{\shortauthors}{O. Malik et al.}

\begin{abstract}
In social sciences, simulating opinion dynamics to study the interplay between homophily and influence, and the subsequent formation of echo chambers, is of great importance. As such, in this paper we investigate echo chambers by implementing a unique social game in which we spawn in a large number of agents, each assigned one of the two opinions on an issue and a finite amount of influence in the form of a game currency. Agents attempt to have an opinion that is a majority at the end of the game, to obtain a reward also paid in the game currency. At the beginning of each round, a randomly selected agent is selected, referred to as a speaker. The second agent is selected in the radius of speaker influence (which is a set subset of the speaker's neighbors) to interact with the speaker as a listener. In this interaction, the speaker proposes a payoff in the game currency from their personal influence budget to persuade the listener to hold the speaker's opinion in future rounds until chosen listener again. The listener can either choose to accept or reject this payoff to hold the speaker's opinion for future rounds. The listener's choice is informed only by their estimate of global majority opinion through a limited view of the opinions of their neighboring agents. We show that the influence game leads to the formation of "echo chambers," or homogeneous clusters of opinions. We also investigate various scenarios to disrupt the creation of such echo chambers, including the introduction of resource disparity between agents with different opinions, initially preferentially assigning opinions to agents, and the introduction of committed agents, who never change their initial opinion.
\end{abstract}



\maketitle

\section{Introduction}

Social simulations have become an important research tool for the social sciences \cite{perc2010coevolutionary, perc2013evolutionary, snijders2010introduction, shirado2013quality}. Combined with the advent of easily collectible, large-scale social datasets, social simulations have enabled research on an unprecedented number of individuals ranging from communities to entire societies. Simulations of social mechanisms and phenomena are often driven by agent-based \cite{Lu2009,raberto2001agent} or network-based simulations \cite{karimi2018homophily,snijders2010introduction}. In such simulations, the agents represented by nodes interact with their neighbors through their connections to model the observed empirical dynamics. One of the most important social processes that is often simulated and captured within these frameworks is opinion dynamics, where individuals make decisions based on their (and their neighbors') personal opinions and/or beliefs. These simulated decisions can range from the dissolution or formation of connections, to the exchange of some incentive, to the alteration of attributes \cite{yuan2018interpretable, pagan2019game}. 

Currently, two core aspects that are used to drive many of these computational models of opinion dynamics are influence and homophily \cite{deffuant2002can} whose interplay decides the outcome of the opinion dynamics for a given social networks. Influence usually represents an ability of one individual to affect opinions of others. It can be modeled as a function of a payoff from a certain influence budget available to the influencer to spreads their opinion to others. Homophily, on the other hand, refers to an agent's proclivity to interact and associate with neighbors that have similar characteristics (i.e., similarity of socio-demographic attributes) and opinions \cite{mcpherson1,reagans}. An empirical example can be found in the emergence of social media, which is widely believed to have facilitated the formation of "filter bubbles" and "echo chambers". Once formed, they limit exposure of individuals within these formations to news that are consistent only with their aligned opinions, and constrains such individuals to only propagating similarly aligned content. These limitations and constraints result in perpetually strengthening homophily at the cost of weakening inter-community ties \cite{sunstein2018legal, toth2021inequality}. The resulting "polarizing" effect has been arising in the politics of many countries recently. Political scientists have contrasted such polarization with the mechanism leading to the "median voter" \cite{downs1957economic}, where two or three credible news sources establish a set of broadly shared core opinions of voters. Meanwhile, increased polarization fueled by social media and driven by homophily and influence within fragmented communities have lead to local consensus and global clustering with distinct separations of clusters from each other \cite{flamino2021shifting,lu2019evolution}. Given the scope of these empirical observations, work on understanding influence and homophily, on the role they play in the formation of echo chambers, and on ways in which the formation of such chambers can be delayed or disrupted has become critical to multiple scientific fields and the preservation of stable democracies around the world \cite{macy2022polarization}.

In this paper we extend the work on simulating the interplay between influence and homophily to investigate a variety of echo chamber models and explore how the changes in the parameter setting of the model change the outcome of this interplay. Following previous approaches
\cite{perc2010coevolutionary,perc2013evolutionary,karimi2018homophily,murase2019structural}, we develop an agent-based model of opinion dynamics. This model is inspired by the naming game model which has become a mathematical archetype for social and behavioral analysis \cite{pickering2016analysis}. 

The basic properties of the naming game were established in \cite{baronchelli2008depth}. Our model follows the binary version of this model but replaces conditions under which an agent changes opinion. In our game agent decisions are made to maximize the probability of the agent holding the majority opinion, estimated from the majority opinion of their neighbors while including a potential payoff from the speaker for holding their opinion. In the binary naming game, agent decisions are made according to the probability of receiving two subsequent interactions with the speakers carrying that same opinion, which also results in a higher probability of that agent having an opinion in agreement with the opinion of the majority of the agent's neighbors in their radius of influence. Then, not surprisingly, the results of the simulation are quite similar for both models. This is illustrated in Fig. \ref{fig:random}, which is similar to Fig. 1 in \cite{zhang2014opinion}, showing the process of merging of initially fine clusters into more coarse ones.  This similarity also extends to the role of committed agents, introduced in \cite{Lu2009}. A novel aspect of our game model is the budget that a speaker can use for payoff of other agents to increase the speaker's chances to hold a majority opinion. Differing this budget to different agents empower them with different influence abilities.

In section 2 we discuss details of our proposed game. In section 3 we present simulated results for this game and investigate various variants of the game to introduce disparities between agents of different opinions and to study the ways in which these disparities affect the consensus formation. Finally, section 4 summarizes and concludes the paper.

\section{Simulation}

An agent $j$ in the game is represented by a node in a graph with a unique label and is initialized with an opinion, $O_j(0)$, and an initial budget in the game currency, $I_j(0)$. Agents that hold the majority opinion at the end of round $T$ gain the reward in the game currency, $I_W$, that increases their budget. The edges in the graph represent pairs of nodes that can directly interact with each other. We assume that the graph has only one component. Each agent $j$ can view the opinions of its $r_k$-th nearest neighbors (where $r_k\leq d$, where $d$ is the diameter of the network) and $M_j(t)$ refers to the local majority opinion (including the agent's own opinion) seen by agent $j$ among all its $r_k$-th nearest neighbors. We refer to $r_k$ as the radius of knowledge, which is the same for all agents. The set of nodes that form the $r_k$-th nearest neighbors of any agent is determined by the topology of the network and remains constant throughout the duration of the game. However, the size of this set may differ depending on the agent's position in the graph. The agents residing in the dense parts of the graph benefit from having their predictions based on the larger number of opinions known to them than to agents residing in the sparse parts of the graph. 

At every round $t$ a speaker, $s$, is randomly selected. The speaker randomly selects a listener, $l$, from its $r_i$-th nearest neighbors (where $r_i$ is referred to as the radius of influence). Both the speaker and listener use a simple forecasting function to predict their expected winnings at the end of the game

\begin{align}
    E(O_j(t), M_j(t), I_j(t)) &= I_W \delta_{O_j(t), M_j(t)} + I_j(t).
\end{align}

In other words, they predict that they will be in the majority opinion if they perceive their opinion to be the same as the majority observed amongst their $r_k$-th nearest neighbors. The interaction can be split into the actions taken by the speaker and the listener.

In the first part of the interaction the speaker determines the amount of game currency to offer to the listener. The speaker does this by calculating the change in their expected winnings if the listener were to hold the speaker's opinion based on the formula

\begin{align}
    I_{s, l} &= 
    \left\{\begin{array}{@{}ll}
      1 & \text{if } E(O_s(t), \tilde{M}_s(t), I_s(t)) > E(O_s(t), M_s(t), I_s(t))  \\
      0 & \text{if } E(O_s(t), \tilde{M}_s(t), I_s(t)) \le E(O_s(t), M_s(t), I_s(t))
\end{array},\right.
\end{align}

where $\tilde{M}_s(t)$ is the local majority opinion seen by the speaker under the assumption that the listener will hold the opinion of the speaker's. The amount $I_{s, l}$ is then transmitted to the listener.

Upon receiving the offer from the speaker, the listener accepts the payoff and commits to holding the speaker's opinion in future rounds with the probability:

\begin{align}
    P_d &= \dfrac{1}{1 + e^{-\Delta I}} ,
\end{align}

where $\Delta I$ represents the expected winnings to the listener upon accepting the speaker's offer, given by

\begin{align}
    \Delta I &= E(O_s(t), \tilde{M}_l(t), I_l(t)) - C_l + I_{s, l} - E(O_s(t), M_l(t), I_l(t)).
\end{align}


As before, $\Tilde{M}_l(t)$ refers to the local majority seen by the listener under the assumption that the listener will hold the speaker's opinion. The cost of changing opinions is $C_l$ and represents reluctance to commit to holding the speaker's opinion, or the extent of social pressure to hold a different opinion than the speaker's. By default we set $C_l = 0$, however committed agents can be modeled within the influence game by setting $C_l = \infty$.

If both agents have the same opinion then we say:

\begin{align}
    E(O_s(t), \tilde{M}_s(t), I_s(t)) &= E(O_s(t), M_s(t), I_s(t)),
\end{align}

In this case, the speaker does not need to spend any game currency so $I_{s, l} = 0$. If the two interacting agents have different opinions, then the speaker must check if spending game currency to convince the listener to switch sides will result in greater winnings. If the speaker's opinion is in the majority by a large margin, it might be the case that even were the listener to switch their opinion it would not change the expected winnings for the speaker and therefore $I_{s, l} = 0$. Similarly, the speaker might have an opinion that is significantly in the minority such that the listener changing their opinion would still not result in a change in winnings for the speaker. Only when the expenditure of game currency will change the majority opinion from the speaker's perspective and will result in a net gain at the end of the game reward will the speaker make a non-zero offer of $I_{s, l} = 1$.

When the listener receives an offer, the probability that it will accept it depends on the expected winnings upon acceptance. Depending on the local majority opinion it perceives, the probability of acceptance may be greater than 0.5 even if the offer made by the speaker is 0. However, the reason for the change is different in this case than in the case when an offer is positive and the reason for acceptance is the influence of the speaker. With the offer being zero, it is the majority of opinions of neighbors within the  radius of knowledge of the listener, in other words homophily, that motivates the listener's acceptance. Using this distinction allows finding what features of the network, incentives, and parameters promote influence over homophily. These findings are of interest for establishing in which social media environments the individuals are prone to follow a few efficient influencers, often referred to as super spreaders \cite{flamino2021shifting}.   

If this interaction is successful, the listener will hold the opinion of the speaker. The listener also gains the payoff offered by the speaker. If the interaction is unsuccessful, nothing happens. Agents can only change their opinion when they are selected as the listener.

While agents are driven by desire to maximize their personal influence in every interaction, they are also pressured to conform to the opinion of their local neighborhood. There are numerous degrees of freedom that can be tuned in this game to alter its dynamics. We will show in the next section how even the simple scenarios lead to interesting results.

\section{Results}
\begin{figure*}[!htp]
	\centering
	\subfigure{\includegraphics[width=0.25\linewidth]{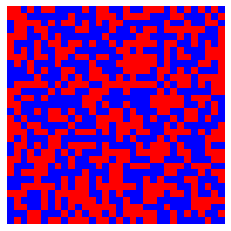}}
	\subfigure{\includegraphics[width=0.25\linewidth]{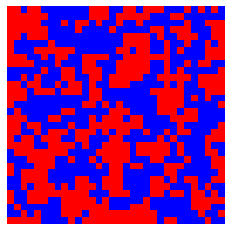}}
	
	\subfigure{\includegraphics[width=0.25\linewidth]{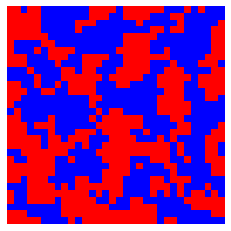}}
	\subfigure{\includegraphics[width=0.25\linewidth]{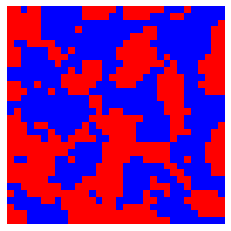}}
	
	\subfigure{\includegraphics[width=0.25\linewidth]{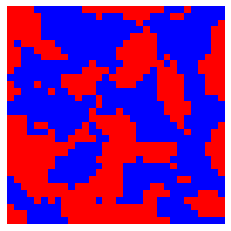}}
	\subfigure{\includegraphics[width=0.25\linewidth]{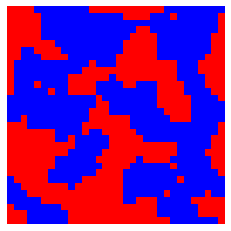}}
	
	\subfigure{\includegraphics[width=0.25\linewidth]{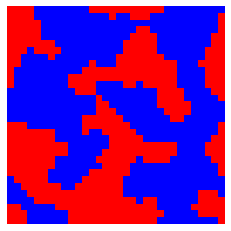}}
	\subfigure{\includegraphics[width=0.25\linewidth]{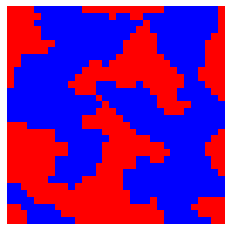}}
	\caption{The progress of the influence game with each agent initialized into one of two randomly selected opinions on a 2d lattice with periodic boundary conditions. Each node is colored by the opinion it holds. The figures show snapshots after every 50 rounds. The heterogeneous starting conditions quickly converge to a few homogeneous clusters. Despite different abstractions of the interplay between influence and homophily in our game and the naming game model, the process of growth of clusters is similar in these models as seen in Fig. 1 in \cite{zhang2014opinion}.} \label{fig:random}
\end{figure*}
\begin{figure}
    \centering
    \includegraphics[width=\linewidth]{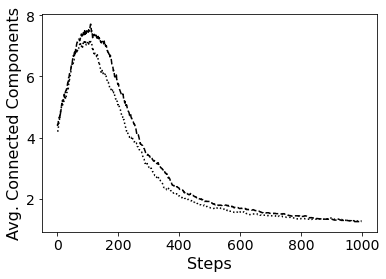}
    \caption{The number of connected components of nodes with the same opinion over time for the derived opinion on a 2d lattice with periodic boundary conditions. The two lines refer to the networks derived from each opinion. As time progresses, the number of components approaches 1, indicating that the opinions are coalescing spatially. }
    \label{fig:comp}
\end{figure}

We populated a 2d lattice with periodic boundary conditions with $30^2$ agents. Each agent can only poll the opinions of and interact with one of its 4 immediate neighbors, thus $r_k = r_i = 1$. We set $ C_j = 1, I_W = 10$ and $I_j(0) = 0$. The fraction of the population having opinion $A$ at round $t$ is denoted by $p_A(t)$. 

The first scenario that we investigated involved random initialization of each node to either of two opinions, labeled $A$ and $B$. As the game proceeds, the disparate agents coalesce into a few large clusters of the same opinion that persist over very long time-scales and no consensus is reached. This effect can be seen in Fig. \ref{fig:random} for a single simulation, and in Fig. \ref{fig:comp}, which shows the average cluster size for both opinions, averaged over 100 simulation runs. The random initialization produces a number of disconnected clusters. As the game progresses the smaller clusters join together into an ever-shrinking number of homogeneous clusters, which frequently converges to one cluster per opinion. This indicates that the dynamics of the influence game lead to community formation without the formation of a global consensus and without underlying community structure in the network topology.

\begin{figure}
	\centering
	\subfigure[]{\label{fig:uneven_init}\includegraphics[width=0.45\linewidth]{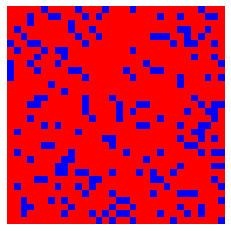}}
	\subfigure[]{\label{fig:uneven_fin}\includegraphics[width=0.45\linewidth]{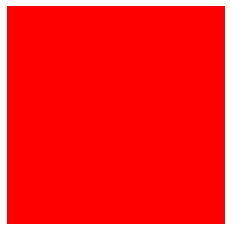}}
	\subfigure[]{\label{fig:droplet_init}\includegraphics[width=0.45\linewidth]{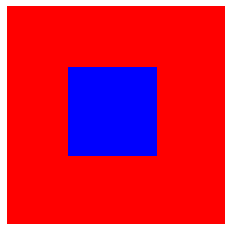}}
	\subfigure[]{\label{fig:droplet_fin}\includegraphics[width=0.45\linewidth]{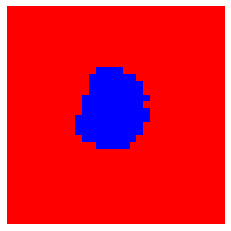}}
	\caption{Results of runs on a 2d lattice with periodic boundary conditions. Each node is colored by the opinion it holds. (a) Random initialization with uneven populations at $t = 0 $. (b) The random initialization at $t = 1000$. The minority opinion has been absorbed. (c) The droplet, or echo chamber, initialization where a minority opinion is initialized in a cluster surrounded by the majority opinion. (d) The droplet configuration at $t=1000$. While the minority opinion has shrunk, it is resilient to being absorbed.}\label{fig:var_init}
\end{figure}

\subsection{Engineering Consensus}

In order to push the game into consensus we investigate different scenarios which can disrupt the homogeneous clusters of opinions that form in the influence game.

\subsubsection*{Larger Radius of Knowledge}

\begin{figure}
    \centering
    \includegraphics[width=\linewidth]{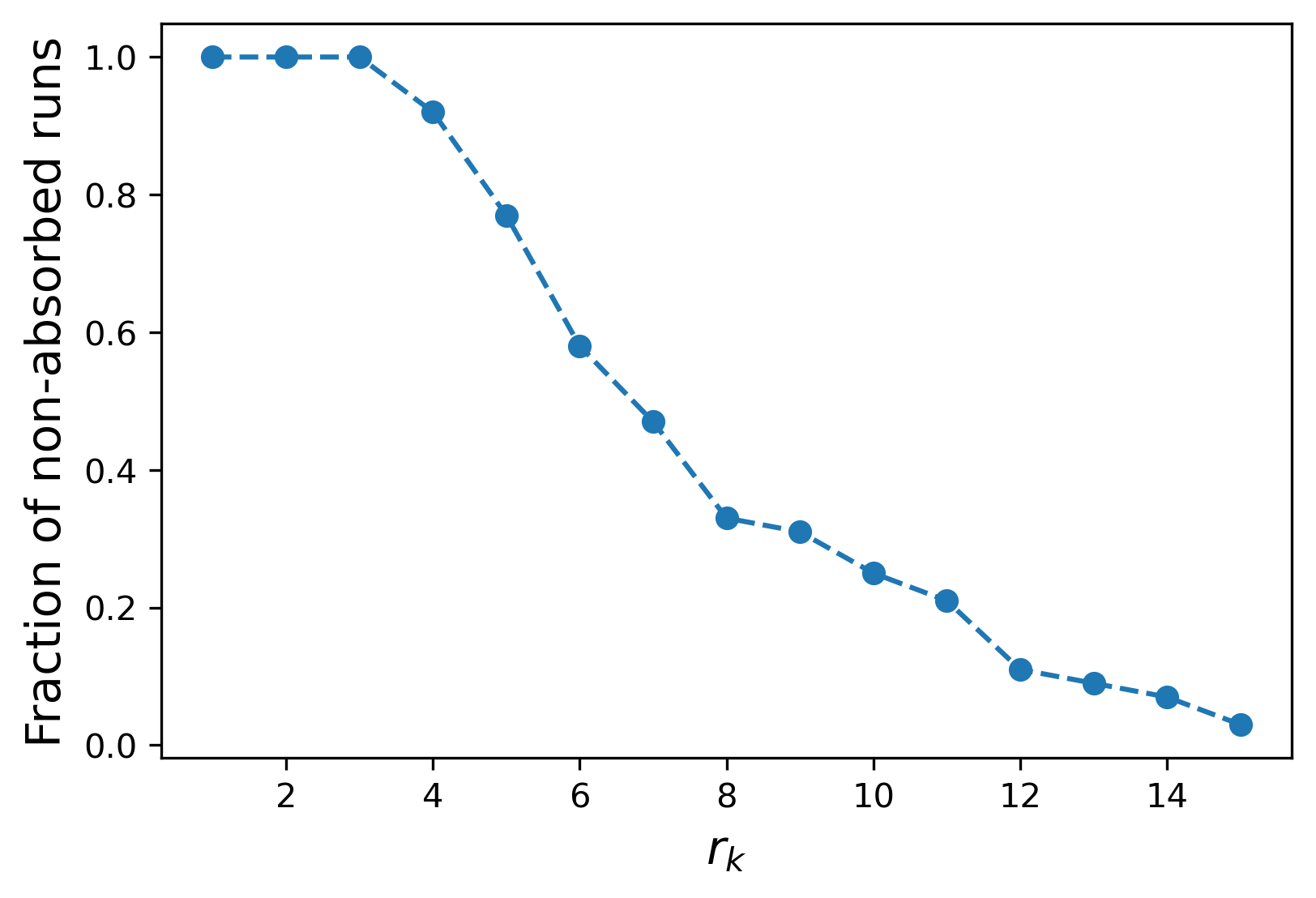}
    \caption{The fraction of non-absorbed runs on a 2d lattice with periodic boundary conditions plotted as a function of the radius of knowledge after $10 * 4$ rounds with $30^2$ agents and $p_A(0) = p_B(0) = 0.5$.}
    \label{fig:abs_var_r_k}
\end{figure}

\begin{figure}
    \centering
    \includegraphics[width=\linewidth]{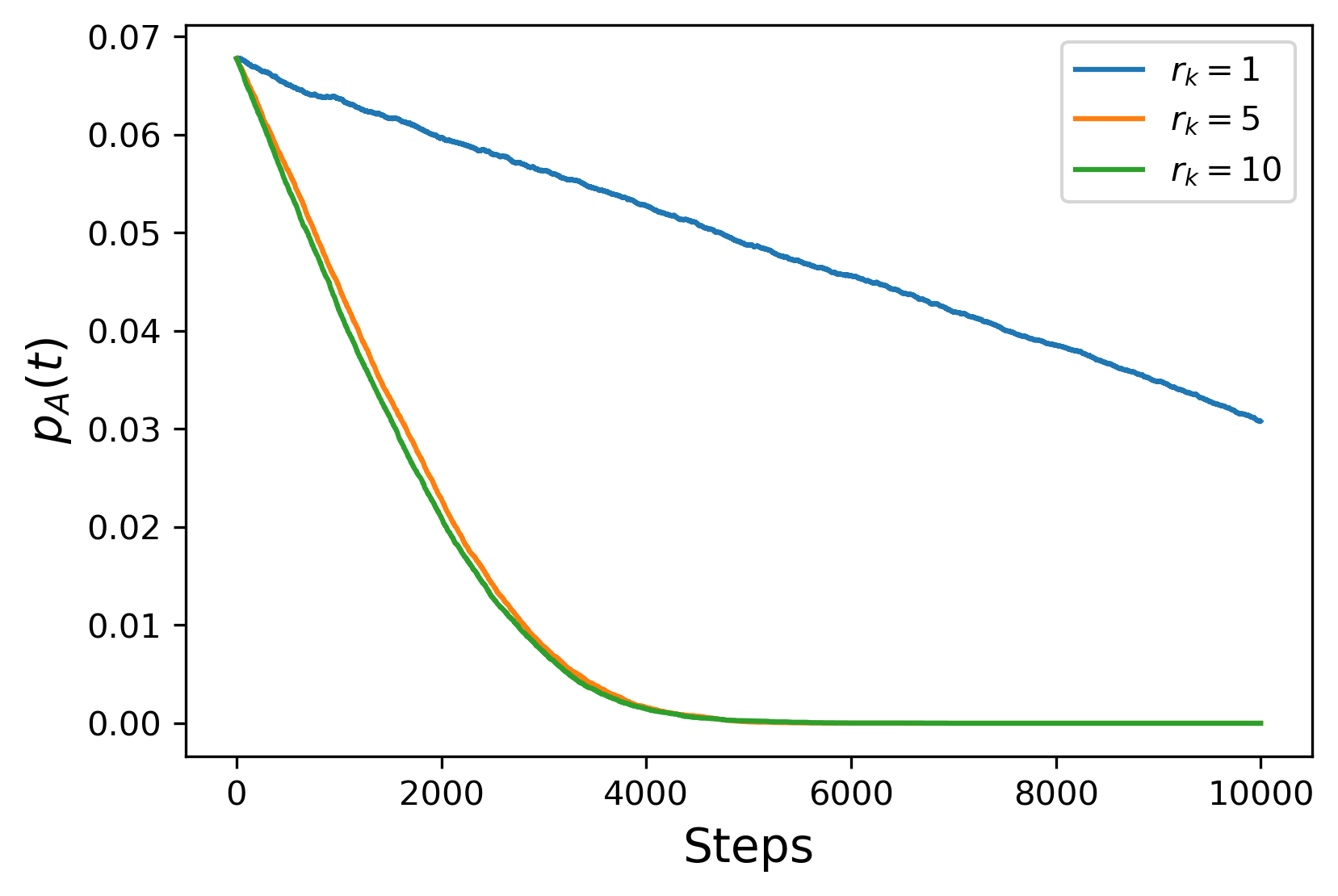}
    \caption{The fraction of agents holding opinion $A$ over time with $30^2$ agents. Opinion $A$ is initialized in the droplet configuration for a 2d lattice with periodic boundary conditions.}
    \label{fig:drop_var_r_k}
\end{figure}

In every interaction, the listener polls all their $r_k$-th nearest neighbors to determine whether or not to accept payoff and hold the speaker's opinion. Even if the listener holds a global minority opinion, but their local majority opinion is not the speaker's opinion, the listener will be unlikely to accept the speaker's payoff. When $r_k = 1$ this leads to the formation of the "echo chambers" shown in Fig.~\ref{fig:comp}. It allows clusters of minority opinions to be resilient against absorption, as shown in Fig.~\ref{fig:var_init}. By increasing $r_k$, the likelihood that an agent will see a dissenting opinion increases. Increasing $r_k$ allows agents to have a more accurate estimate of the global majority opinion. It is easy to predict that in scale-free networks \cite{onnela2007structure}, the number of semi-stable clusters will be equal to the number of hubs in the network, even if  $r_k=1$. In social networks with strong community structures, the semi-stable number of clusters will likely be defined by the number of communities.    

Fig.~\ref{fig:abs_var_r_k} shows how changing the radius of knowledge leads to an increased likelihood of consensus formation. Increasing the radius of knowledge also leads to the dissolution of the minority opinion in the droplet configuration, as shown in Fig.~\ref{fig:drop_var_r_k}.

\subsubsection*{Resource Disparity}

\begin{figure}
    \centering
    \includegraphics[width=\linewidth]{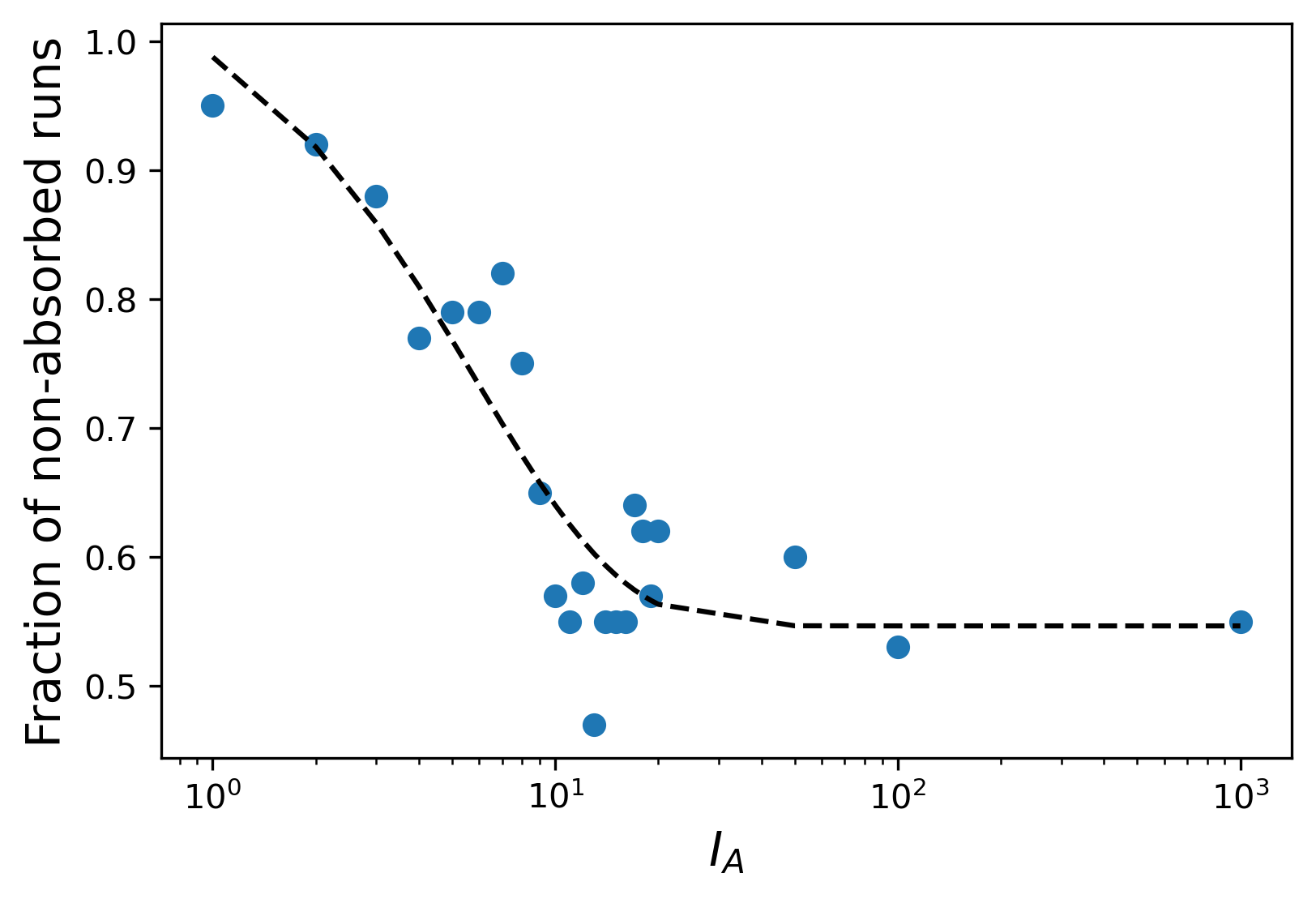}
    \caption{The fraction of runs on a 2d lattice with periodic boundary conditions that do not reach consensus as a function of the game currency incentive offered by speakers holding opinion A. The dotted line shows an exponential fit.}
    \label{fig:rsrc_disp}
\end{figure}

We modify our game so that one of the two opinions (labeled A and B) has a resource advantage. If the speaker holds opinion A, they are able to offer an influence incentive in the amount of $I_A$ in game currency greater than 1, but when the speaker holds opinion B, they are able to offer an influence incentive at the level of one unit of the game currency. We randomly initialize both opinions with an equal number of agents and perform 100 runs for each value of $I_A$ and determine the number of runs that reach consensus at the end of $5*10^4$ rounds. The results are plotted in Fig.~\ref{fig:rsrc_disp}. As we can see, when $I_A = 1$ we do not observe any run reaching consensus. However, as $I_A$ begins to increase the probability that a run will not reach consensus decays exponentially. Yet, the probability levels out at a non-zero value for $I_A > I_W$ (with $I_W = 10$). Initially, as $I_A$ increases, the probability that a listener will hold opinion A increases. However, this probability quickly saturates when $I_A > I_W$.

\subsubsection*{Topological Disparity}

\begin{figure}
    \centering
    \includegraphics[width=\linewidth]{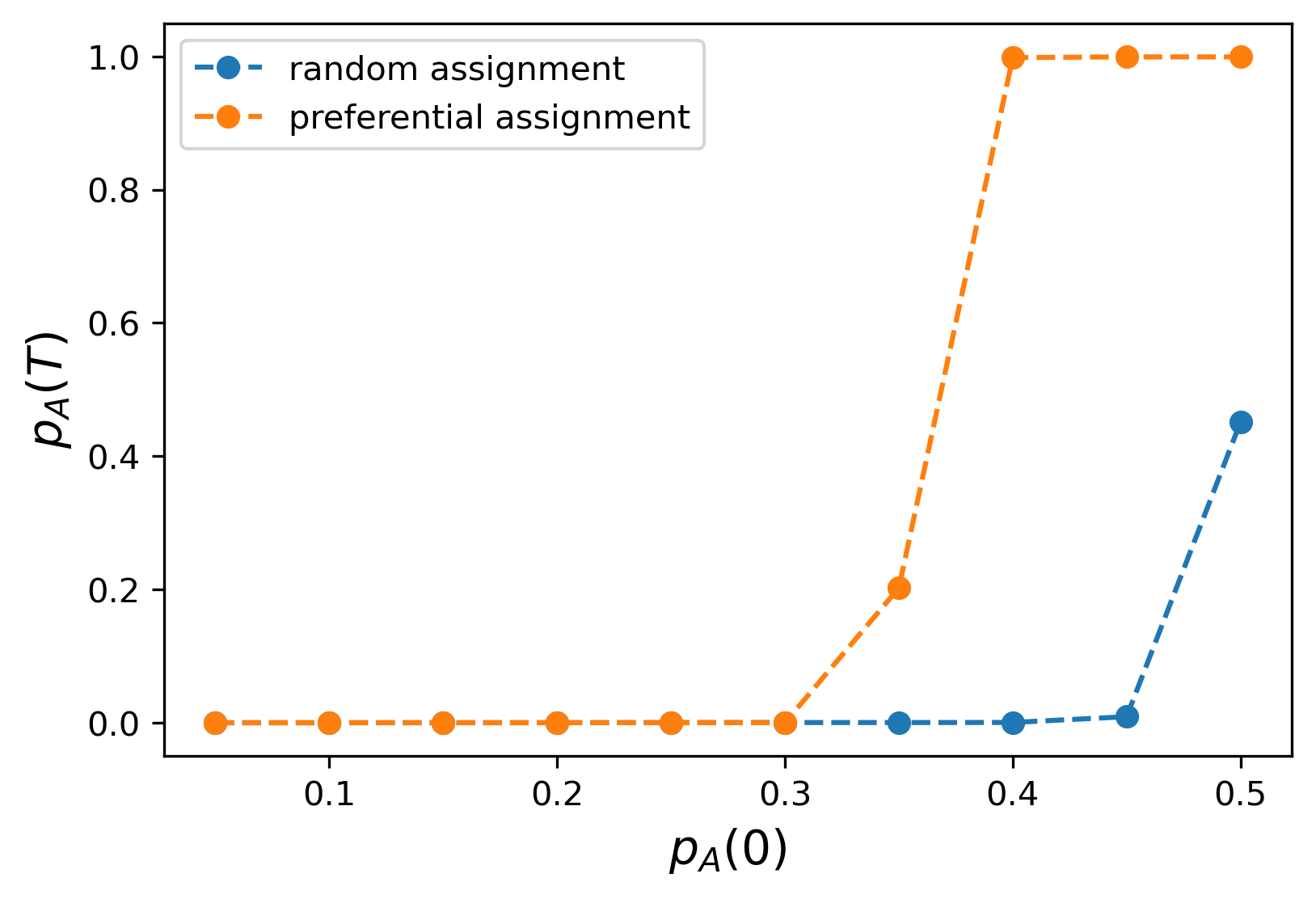}
    \caption{The fraction of nodes with opinion A at the termination of the game plotted against the initial fraction of nodes holding opinion A when the game runs on a 2d lattice with periodic boundary conditions.}
    \label{fig:top_disp}
\end{figure}

In order to model agents of one opinion having greater reach than those of the other opinion, we spawn $1000$ agents on a Barab\'{a}si--Albert (BA) network with coordination number $m = 4$. We traverse the node-list of the graph in descending order of node degree and assign these nodes the opinion A until we reach the designated number of agents. The remaining nodes are assigned the opinion B. We vary the number of agents holding opinion A and generate 100 network instances for each initial value. We run the game for $5 * 10^4$ rounds and determine the average number of agents with opinion A at the termination of the game. We can see from Fig.~\ref{fig:top_disp} that when nodes are assigned randomly even a small amount of disparity in the populations of the two opinions causes the minority opinion to be absorbed. On the other hand, when opinion A is assigned preferentially, it can lead to consensus even when opinion A is in the minority in agreement with the outcomes of running a naming game with variable commitment \cite{Niu2017}.

\subsection{Non-binary Opinion Dynamics}

\begin{figure*}[!htp]
	\centering
	\subfigure{\includegraphics[width=0.25\linewidth]{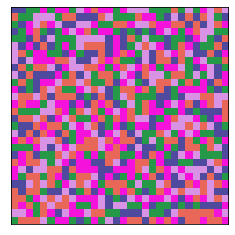}}
	\subfigure{\includegraphics[width=0.25\linewidth]{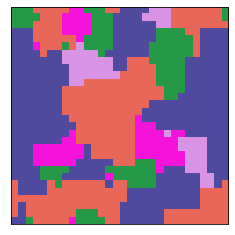}}
	
	\subfigure{\includegraphics[width=0.25\linewidth]{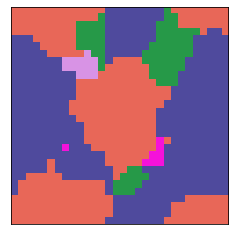}}
	\subfigure{\includegraphics[width=0.25\linewidth]{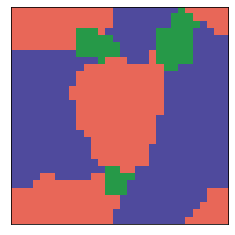}}
	
	\subfigure{\includegraphics[width=0.25\linewidth]{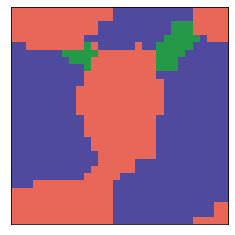}}
	\subfigure{\includegraphics[width=0.25\linewidth]{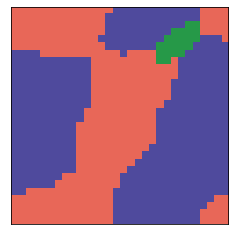}}
	
	\caption{The progress of the influence game on a 2d lattice with periodic boundary conditions with each agent initialized into one of five randomly selected opinions. The figures show snapshots after every 10000 rounds. Some opinions are absorbed, but consensus is not reached.}\label{fig:multi_random}
\end{figure*}

\begin{figure}
    \centering
    \includegraphics[width=\linewidth]{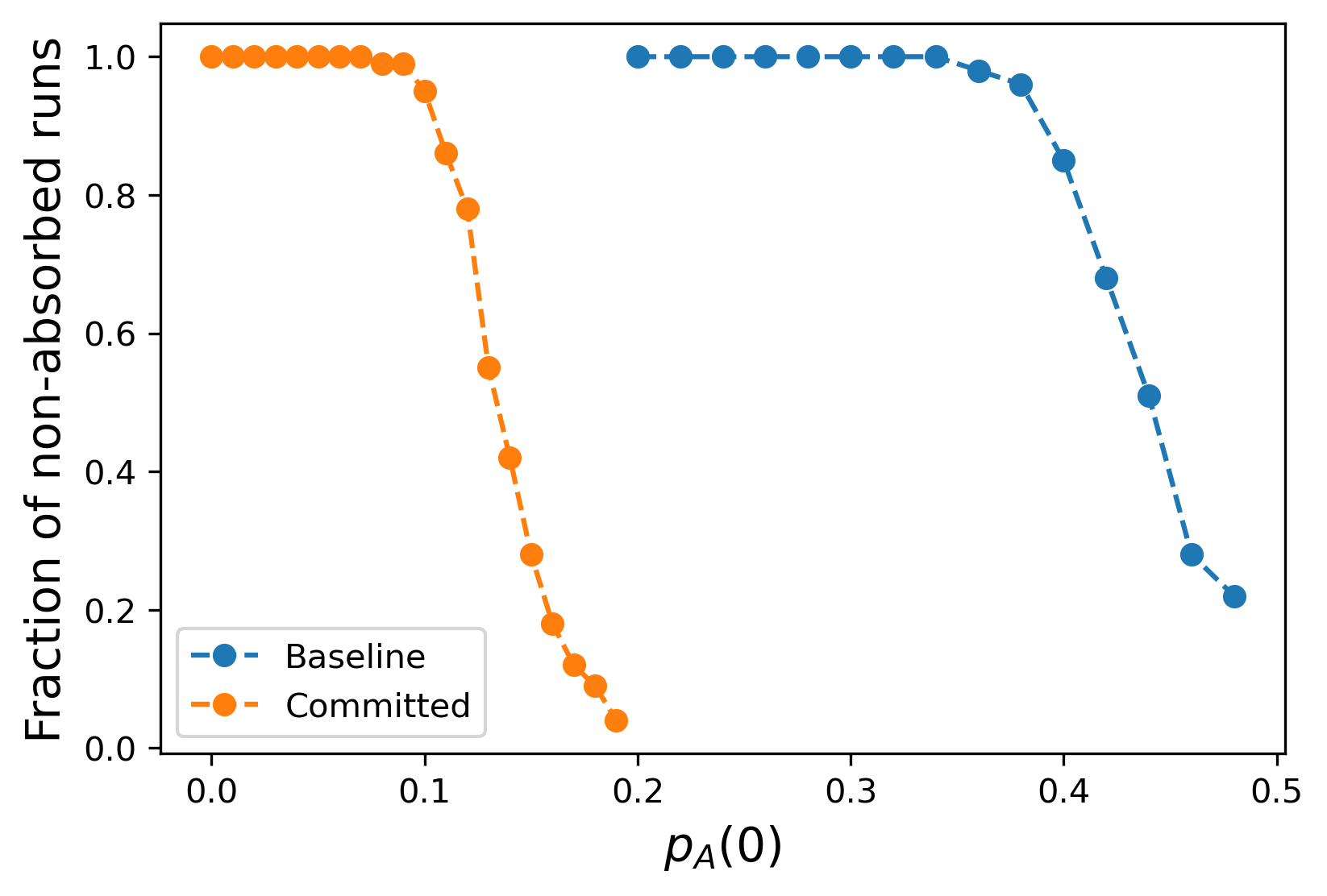}
    \caption{The fraction of non-absorbing runs on a 2d lattice with periodic boundary conditions as a function of the initial fraction of nodes with opinion A, with four other competing opinions. The blue line shows opinion A without any committed agents, while the orange lies shows the initial population of opinion A composed of committed agents.}
    \label{fig:multi_comm}
\end{figure}

We also investigate the scenario when there are more than two competing opinions. We randomly initialize $30^2$ agents into one of five opinions on a 2d grid with periodic boundary conditions and run the influence game for 5000 rounds. We can see from Fig.~\ref{fig:multi_random} that while several of the opinions are absorbed, we do not see consensus formation. Instead, similar to the scenario with two competing opinions, we see the formation of stable, homogeneous clusters of opinions, with no total absorption achieved.

\subsubsection*{Impact of Committed Agents} A committed agent is one that never changes an opinion with which it was initialized in~\cite{Lu2009}. In our framework, we can realize committed agents by setting $C_j = \infty$. We populate one of the five opinions, including opinion A, with committed agents and vary the initial population of nodes with this opinion, while having the remaining nodes equally split among the other four opinions. From Fig~\ref{fig:multi_comm} we can see that in the absence of committed agents, an opinion needs to be in significant majority, approximately 0.44 of the total population, for the probability of reaching consensus to be 0.5. This can be contrasted with the case where opinion A is initialized with committed agents and an initial population of just 0.15 is enough for the probability of reaching consensus to be at least 0.5.

\section{Conclusion}

In this paper we present a model that aims to capture the spread of opinions through social pressure where every agent is concerned with maximizing their own game currency budget. We designed this game with inspiration from the simplest version of the naming game, which allows only binary opinions. Despite the different abstractions of the interplay between homophily and influence used in these two models, described in the section Introduction, we observe the similar outcomes of the opinion dynamics in these models. We observe that randomly initializing every agent with either of the opinions leads to consensus if one opinion is held by significantly fewer agents. On the other hand, we get meta-stable, homogeneous communities, or echo chambers, if the two opinions have roughly equal populations. We also see that if we initialize a minority opinion as a homogeneous community surrounded by a different opinion it is remarkably resilient to being absorbed. This is because each agent has access only to local information, agents embedded in homogeneous communities do not feel the same pressure to conform to the majority opinion as agents holding minority opinions in more heterogeneous environments. Our model therefore captures two distinct phenomena. The first is the formation of echo chambers as agents with the same opinions begin to coalesce, forming homogeneous communities from heterogeneous starting conditions. We believe this is caused by the behavior of individuals preferring to associate with those who share their opinions. Secondly, we showed how homogeneous minority communities are resilient to pressure from their environment to conform. These echo chambers reinforce the minority opinion among its agents, and we believe this captures the real world behavior of tight-knit communities that preserve their distinct practices, whether religious, political or cultural, despite being surrounded by often hostile opposition.

Our model also captures how the introduction of various disparities between opinions can disrupt these echo chambers. If agents having one opinion are able to offer a greater incentive in interactions than agents of other opinions, the likelihood of reaching consensus increases. This scenario can be seen as representing a concerted effort, such as an educational campaign or an advertisement campaign, to spread a particular opinion among a population. 

We also showed that increasing the radius of knowledge can lead to the disruption of echo chambers. As agents have access to opinions from a broader range of the population, they are able to more accurately estimate the global consensus. As the radius of knowledge increases, the likelihood of global consensus sharply increases. An increased radius of knowledge also leads to echo chambers of minority opinion being quickly resolved. By contrast, these echo chambers are remarkably resilient when the radius of knowledge is small. These results can be compared with the waning commitment model of~\cite{Niu2017}. In the waning commitment model, agents become open to the possibility of accepting opposing opinion after having $w$ serial interactions with agents of holding such opinion. In the influence game, the radius of knowledge controls the likelihood of an agent seeing more dissenting opinions. Furthermore, similar phenomena can be found in empirical studies, where partisan individuals taken outside of their previously established echo chambers (specifically dedicated Fox News viewers being externally incentivized to watch CNN) consistently began to identify and admit to the biased content filtering effects of the echo chambers they were a part of~\cite{broockman2022manifold}.

We also study how minority opinions can become dominant within a population within the framework of our model. Preferentially assigning one opinion to nodes with a higher degree can lead to consensus, even when the advantaged opinion is the minority, with a minority opinion fraction threshold of approximately $40 \%$. We also investigated the scenario of more than two competing opinions, and saw that this case is particularly unlikely to reach consensus, even when one opinion is in significant plurality. In contrast, the introduction of a small number of committed agents, approximately $20 \%$ of the total population, can lead to consensus. We note that our values of this tipping point are higher than those associated with the naming game with committed agents, which is approximately $10 \%$~\cite{Xie_2011}. However, a recent empirical study where individuals attempt to reach consensus for naming an object in a series of pairwise interactions, with pairwise (but not global) consensus being incentivized, suggest that the critical threshold of committed agents required to achieve global consensus lies between $20 \%$ and $30 \%$, agreeing more with our model~\cite{Baronchelli2018}. Similarly, the critical threshold is approximately $35 \%$ for small values of $w$ in the waning commitment model~\cite{Niu2017}. 

In future work we plan to introduce agents that have memory and that are able to adapt their strategies as the game proceeds. Furthermore, our game can be modified so that at every interaction the speaker loses the amount of the game currency that they offer to the speaker, and the amount that they offer can be varied. This increases the variety of strategies that an agent can adopt and it would be interesting to see if a dominant \balance  strategy emerges.

\bibliographystyle{ACM-Reference-Format}
\bibliography{bibliography}


\begin{thebibliography}{27}


\ifx \showCODEN    \undefined \def \showCODEN     #1{\unskip}     \fi
\ifx \showDOI      \undefined \def \showDOI       #1{#1}\fi
\ifx \showISBNx    \undefined \def \showISBNx     #1{\unskip}     \fi
\ifx \showISBNxiii \undefined \def \showISBNxiii  #1{\unskip}     \fi
\ifx \showISSN     \undefined \def \showISSN      #1{\unskip}     \fi
\ifx \showLCCN     \undefined \def \showLCCN      #1{\unskip}     \fi
\ifx \shownote     \undefined \def \shownote      #1{#1}          \fi
\ifx \showarticletitle \undefined \def \showarticletitle #1{#1}   \fi
\ifx \showURL      \undefined \def \showURL       {\relax}        \fi
\providecommand\bibfield[2]{#2}
\providecommand\bibinfo[2]{#2}
\providecommand\natexlab[1]{#1}
\providecommand\showeprint[2][]{arXiv:#2}

\bibitem[\protect\citeauthoryear{Baronchelli, Loreto, and Steels}{Baronchelli
  et~al\mbox{.}}{2008}]%
        {baronchelli2008depth}
\bibfield{author}{\bibinfo{person}{Andrea Baronchelli},
  \bibinfo{person}{Vittorio Loreto}, {and} \bibinfo{person}{Luc Steels}.}
  \bibinfo{year}{2008}\natexlab{}.
\newblock \showarticletitle{In-depth analysis of the Naming Game dynamics: the
  homogeneous mixing case}.
\newblock \bibinfo{journal}{\emph{International Journal of Modern Physics C}}
  \bibinfo{volume}{19}, \bibinfo{number}{05} (\bibinfo{year}{2008}),
  \bibinfo{pages}{785--812}.
\newblock
\urldef\tempurl%
\url{https://doi.org/10.1142/S0129183108012522}
\showURL{%
\tempurl}


\bibitem[\protect\citeauthoryear{Broockman and Kalla}{Broockman and
  Kalla}{2022}]%
        {broockman2022manifold}
\bibfield{author}{\bibinfo{person}{David Broockman} {and}
  \bibinfo{person}{Joshua Kalla}.} \bibinfo{year}{2022}\natexlab{}.
\newblock \showarticletitle{The manifold effects of partisan media on
  viewers’ beliefs and attitudes: A field experiment with Fox News viewers}.
\newblock \bibinfo{journal}{\emph{OSF Preprints}}  \bibinfo{volume}{1}
  (\bibinfo{year}{2022}), \bibinfo{pages}{1--42}.
\newblock


\bibitem[\protect\citeauthoryear{Centola, Becker, Brackbill, and
  Baronchelli}{Centola et~al\mbox{.}}{2018}]%
        {Baronchelli2018}
\bibfield{author}{\bibinfo{person}{Damon Centola}, \bibinfo{person}{Joshua
  Becker}, \bibinfo{person}{Devon Brackbill}, {and} \bibinfo{person}{Andrea
  Baronchelli}.} \bibinfo{year}{2018}\natexlab{}.
\newblock \showarticletitle{Experimental evidence for tipping points in social
  convention}.
\newblock \bibinfo{journal}{\emph{Science}} \bibinfo{volume}{360},
  \bibinfo{number}{6393} (\bibinfo{year}{2018}), \bibinfo{pages}{1116--1119}.
\newblock
\urldef\tempurl%
\url{https://doi.org/10.1126/science.aas8827}
\showURL{%
\tempurl}


\bibitem[\protect\citeauthoryear{Deffuant, Amblard, Weisbuch, and
  Faure}{Deffuant et~al\mbox{.}}{2002}]%
        {deffuant2002can}
\bibfield{author}{\bibinfo{person}{Guillaume Deffuant},
  \bibinfo{person}{Fr{\'e}d{\'e}ric Amblard}, \bibinfo{person}{G{\'e}rard
  Weisbuch}, {and} \bibinfo{person}{Thierry Faure}.}
  \bibinfo{year}{2002}\natexlab{}.
\newblock \showarticletitle{How can extremism prevail? A study based on the
  relative agreement interaction model}.
\newblock \bibinfo{journal}{\emph{Journal of artificial societies and social
  simulation}} \bibinfo{volume}{5}, \bibinfo{number}{4} (\bibinfo{year}{2002}).
\newblock
\urldef\tempurl%
\url{https://www.jstor.org/stable/20868899}
\showURL{%
\tempurl}


\bibitem[\protect\citeauthoryear{Downs et~al\mbox{.}}{Downs
  et~al\mbox{.}}{1957}]%
        {downs1957economic}
\bibfield{author}{\bibinfo{person}{Anthony Downs} {et~al\mbox{.}}}
  \bibinfo{year}{1957}\natexlab{}.
\newblock \bibinfo{booktitle}{\emph{An economic theory of democracy}}.
\newblock \bibinfo{publisher}{Harper}, \bibinfo{address}{New York, NY, USA}.
\newblock


\bibitem[\protect\citeauthoryear{Flamino, Galezzi, Feldman, Macy, Cross, Zhou,
  Serafino, Bovet, Makse, and Szymanski}{Flamino et~al\mbox{.}}{2021}]%
        {flamino2021shifting}
\bibfield{author}{\bibinfo{person}{James Flamino}, \bibinfo{person}{Alessandro
  Galezzi}, \bibinfo{person}{Stuart Feldman}, \bibinfo{person}{Michael~W Macy},
  \bibinfo{person}{Brendan Cross}, \bibinfo{person}{Zhenkun Zhou},
  \bibinfo{person}{Matteo Serafino}, \bibinfo{person}{Alexandre Bovet},
  \bibinfo{person}{Hernan~A Makse}, {and} \bibinfo{person}{Boleslaw~K
  Szymanski}.} \bibinfo{year}{2021}\natexlab{}.
\newblock \showarticletitle{Shifting Polarization and Twitter News Influencers
  between two US Presidential Elections}.
\newblock \bibinfo{journal}{\emph{arXiv preprint}} \bibinfo{volume}{2001},
  \bibinfo{number}{02505} (\bibinfo{year}{2021}), \bibinfo{pages}{1--41}.
\newblock
\urldef\tempurl%
\url{https://doi.org/10.48550/arXiv.2111.02505}
\showURL{%
\tempurl}


\bibitem[\protect\citeauthoryear{Karimi, G{\'e}nois, Wagner, Singer, and
  Strohmaier}{Karimi et~al\mbox{.}}{2018}]%
        {karimi2018homophily}
\bibfield{author}{\bibinfo{person}{Fariba Karimi}, \bibinfo{person}{Mathieu
  G{\'e}nois}, \bibinfo{person}{Claudia Wagner}, \bibinfo{person}{Philipp
  Singer}, {and} \bibinfo{person}{Markus Strohmaier}.}
  \bibinfo{year}{2018}\natexlab{}.
\newblock \showarticletitle{Homophily influences ranking of minorities in
  social networks}.
\newblock \bibinfo{journal}{\emph{Scientific reports}} \bibinfo{volume}{8},
  \bibinfo{number}{1} (\bibinfo{year}{2018}), \bibinfo{pages}{1--12}.
\newblock
\urldef\tempurl%
\url{https://doi.org/10.1038/s41598-018-29405-7}
\showURL{%
\tempurl}


\bibitem[\protect\citeauthoryear{Lu, Korniss, and Szymanski}{Lu
  et~al\mbox{.}}{2009}]%
        {Lu2009}
\bibfield{author}{\bibinfo{person}{Qiming Lu}, \bibinfo{person}{G. Korniss},
  {and} \bibinfo{person}{Boleslaw~K. Szymanski}.}
  \bibinfo{year}{2009}\natexlab{}.
\newblock \showarticletitle{The Naming Game in social networks: community
  formation and consensus engineering}.
\newblock \bibinfo{journal}{\emph{Journal of Economic Interaction and
  Coordination}} \bibinfo{volume}{4}, \bibinfo{number}{2} (\bibinfo{date}{29
  Jul} \bibinfo{year}{2009}), \bibinfo{pages}{221}.
\newblock
\showISSN{1860-7128}
\urldef\tempurl%
\url{https://doi.org/10.1007/s11403-009-0057-7}
\showURL{%
\tempurl}


\bibitem[\protect\citeauthoryear{Lu, Gao, and Szymanski}{Lu
  et~al\mbox{.}}{2019}]%
        {lu2019evolution}
\bibfield{author}{\bibinfo{person}{Xiaoyan Lu}, \bibinfo{person}{Jianxi Gao},
  {and} \bibinfo{person}{Boleslaw~K Szymanski}.}
  \bibinfo{year}{2019}\natexlab{}.
\newblock \showarticletitle{The evolution of polarization in the legislative
  branch of government}.
\newblock \bibinfo{journal}{\emph{Journal of the Royal Society Interface}}
  \bibinfo{volume}{16}, \bibinfo{number}{156} (\bibinfo{year}{2019}),
  \bibinfo{pages}{20190010}.
\newblock
\urldef\tempurl%
\url{https://doi.org/10.1098/rsif.2019.0010}
\showURL{%
\tempurl}


\bibitem[\protect\citeauthoryear{Macy, Ma, Tabin, Gao, and Szymanski}{Macy
  et~al\mbox{.}}{2021}]%
        {macy2022polarization}
\bibfield{author}{\bibinfo{person}{Michael~W Macy}, \bibinfo{person}{Manqing
  Ma}, \bibinfo{person}{Daniel~R Tabin}, \bibinfo{person}{Jianxi Gao}, {and}
  \bibinfo{person}{Boleslaw~K Szymanski}.} \bibinfo{year}{2021}\natexlab{}.
\newblock \showarticletitle{Polarization and Tipping Points}.
\newblock \bibinfo{journal}{\emph{Proceedings of the National Academy of
  Sciences}}  \bibinfo{volume}{118} (\bibinfo{year}{2021}),
  \bibinfo{pages}{e2102144118}.
\newblock
Issue 50.
\urldef\tempurl%
\url{https://doi.org/10.1073/pnas.2102144118}
\showURL{%
\tempurl}


\bibitem[\protect\citeauthoryear{McPherson, Smith-Lovin, and Cook}{McPherson
  et~al\mbox{.}}{2001}]%
        {mcpherson1}
\bibfield{author}{\bibinfo{person}{Miller McPherson}, \bibinfo{person}{Lynn
  Smith-Lovin}, {and} \bibinfo{person}{James~M Cook}.}
  \bibinfo{year}{2001}\natexlab{}.
\newblock \showarticletitle{Birds of a feather: Homophily in social networks}.
\newblock \bibinfo{journal}{\emph{Annu. Rev. Sociol.}} \bibinfo{volume}{27},
  \bibinfo{number}{1} (\bibinfo{year}{2001}), \bibinfo{pages}{415--444}.
\newblock
\urldef\tempurl%
\url{https://doi.org/10.1146/annurev.soc.27.1.415}
\showURL{%
\tempurl}


\bibitem[\protect\citeauthoryear{Murase, Jo, T{\"o}r{\"o}k, Kert{\'e}sz, and
  Kaski}{Murase et~al\mbox{.}}{2019}]%
        {murase2019structural}
\bibfield{author}{\bibinfo{person}{Yohsuke Murase}, \bibinfo{person}{Hang-Hyun
  Jo}, \bibinfo{person}{J{\'a}nos T{\"o}r{\"o}k}, \bibinfo{person}{J{\'a}nos
  Kert{\'e}sz}, {and} \bibinfo{person}{Kimmo Kaski}.}
  \bibinfo{year}{2019}\natexlab{}.
\newblock \showarticletitle{Structural transition in social networks: The role
  of homophily}.
\newblock \bibinfo{journal}{\emph{Scientific reports}} \bibinfo{volume}{9},
  \bibinfo{number}{1} (\bibinfo{year}{2019}), \bibinfo{pages}{1--8}.
\newblock
\urldef\tempurl%
\url{https://doi.org/10.1038/s41598-019-40990-z}
\showURL{%
\tempurl}


\bibitem[\protect\citeauthoryear{Niu, Doyle, Korniss, and Szymanski}{Niu
  et~al\mbox{.}}{2017}]%
        {Niu2017}
\bibfield{author}{\bibinfo{person}{Xiang Niu}, \bibinfo{person}{Casey Doyle},
  \bibinfo{person}{Gyorgy Korniss}, {and} \bibinfo{person}{Boleslaw~K.
  Szymanski}.} \bibinfo{year}{2017}\natexlab{}.
\newblock \showarticletitle{The impact of variable commitment in the Naming
  Game on consensus formation}.
\newblock \bibinfo{journal}{\emph{Scientific Reports}} \bibinfo{volume}{7},
  \bibinfo{number}{1} (\bibinfo{date}{02 Feb} \bibinfo{year}{2017}),
  \bibinfo{pages}{41750}.
\newblock
\showISSN{2045-2322}
\urldef\tempurl%
\url{https://doi.org/10.1038/srep41750}
\showURL{%
\tempurl}


\bibitem[\protect\citeauthoryear{Onnela, Saram{\"a}ki, Hyv{\"o}nen, Szab{\'o},
  Lazer, Kaski, Kert{\'e}sz, and Barab{\'a}si}{Onnela et~al\mbox{.}}{2007}]%
        {onnela2007structure}
\bibfield{author}{\bibinfo{person}{J-P Onnela}, \bibinfo{person}{Jari
  Saram{\"a}ki}, \bibinfo{person}{Jorkki Hyv{\"o}nen},
  \bibinfo{person}{Gy{\"o}rgy Szab{\'o}}, \bibinfo{person}{David Lazer},
  \bibinfo{person}{Kimmo Kaski}, \bibinfo{person}{J{\'a}nos Kert{\'e}sz}, {and}
  \bibinfo{person}{A-L Barab{\'a}si}.} \bibinfo{year}{2007}\natexlab{}.
\newblock \showarticletitle{Structure and tie strengths in mobile communication
  networks}.
\newblock \bibinfo{journal}{\emph{Proceedings of the National Academy of
  sciences}} \bibinfo{volume}{104}, \bibinfo{number}{18}
  (\bibinfo{year}{2007}), \bibinfo{pages}{7332--7336}.
\newblock
\urldef\tempurl%
\url{https://doi.org/10.1073/pnas.0610245104}
\showURL{%
\tempurl}


\bibitem[\protect\citeauthoryear{Pagan and D{\"o}rfler}{Pagan and
  D{\"o}rfler}{2019}]%
        {pagan2019game}
\bibfield{author}{\bibinfo{person}{Nicol{\`o} Pagan} {and}
  \bibinfo{person}{Florian D{\"o}rfler}.} \bibinfo{year}{2019}\natexlab{}.
\newblock \showarticletitle{Game theoretical inference of human behavior in
  social networks}.
\newblock \bibinfo{journal}{\emph{Nature communications}} \bibinfo{volume}{10},
  \bibinfo{number}{1} (\bibinfo{year}{2019}), \bibinfo{pages}{1--12}.
\newblock
\urldef\tempurl%
\url{https://doi.org/10.1038/s41467-019-13148-8}
\showURL{%
\tempurl}


\bibitem[\protect\citeauthoryear{Perc, G{\'o}mez-Gardenes, Szolnoki,
  Flor{\'\i}a, and Moreno}{Perc et~al\mbox{.}}{2013}]%
        {perc2013evolutionary}
\bibfield{author}{\bibinfo{person}{Matja{\v{z}} Perc},
  \bibinfo{person}{Jes{\'u}s G{\'o}mez-Gardenes}, \bibinfo{person}{Attila
  Szolnoki}, \bibinfo{person}{Luis~M Flor{\'\i}a}, {and} \bibinfo{person}{Yamir
  Moreno}.} \bibinfo{year}{2013}\natexlab{}.
\newblock \showarticletitle{Evolutionary dynamics of group interactions on
  structured populations: a review}.
\newblock \bibinfo{journal}{\emph{Journal of the royal society interface}}
  \bibinfo{volume}{10}, \bibinfo{number}{80} (\bibinfo{year}{2013}),
  \bibinfo{pages}{20120997}.
\newblock
\urldef\tempurl%
\url{https://doi.org/10.1098/rsif.2012.0997}
\showURL{%
\tempurl}


\bibitem[\protect\citeauthoryear{Perc and Szolnoki}{Perc and Szolnoki}{2010}]%
        {perc2010coevolutionary}
\bibfield{author}{\bibinfo{person}{Matja{\v{z}} Perc} {and}
  \bibinfo{person}{Attila Szolnoki}.} \bibinfo{year}{2010}\natexlab{}.
\newblock \showarticletitle{Coevolutionary games—a mini review}.
\newblock \bibinfo{journal}{\emph{BioSystems}} \bibinfo{volume}{99},
  \bibinfo{number}{2} (\bibinfo{year}{2010}), \bibinfo{pages}{109--125}.
\newblock
\urldef\tempurl%
\url{https://doi.org/10.1016/j.biosystems.2009.10.003}
\showURL{%
\tempurl}


\bibitem[\protect\citeauthoryear{Pickering, Szymanski, and Lim}{Pickering
  et~al\mbox{.}}{2016}]%
        {pickering2016analysis}
\bibfield{author}{\bibinfo{person}{William Pickering},
  \bibinfo{person}{Boleslaw~K Szymanski}, {and} \bibinfo{person}{Chjan Lim}.}
  \bibinfo{year}{2016}\natexlab{}.
\newblock \showarticletitle{Analysis of the high-dimensional naming game with
  committed minorities}.
\newblock \bibinfo{journal}{\emph{Physical Review E}} \bibinfo{volume}{93},
  \bibinfo{number}{5} (\bibinfo{year}{2016}), \bibinfo{pages}{052311}.
\newblock
\urldef\tempurl%
\url{http://doi.org/10.1103/PhysRevE.93.052311}
\showURL{%
\tempurl}


\bibitem[\protect\citeauthoryear{Raberto, Cincotti, Focardi, and
  Marchesi}{Raberto et~al\mbox{.}}{2001}]%
        {raberto2001agent}
\bibfield{author}{\bibinfo{person}{Marco Raberto}, \bibinfo{person}{Silvano
  Cincotti}, \bibinfo{person}{Sergio~M Focardi}, {and} \bibinfo{person}{Michele
  Marchesi}.} \bibinfo{year}{2001}\natexlab{}.
\newblock \showarticletitle{Agent-based simulation of a financial market}.
\newblock \bibinfo{journal}{\emph{Physica A: Statistical Mechanics and its
  Applications}} \bibinfo{volume}{299}, \bibinfo{number}{1-2}
  (\bibinfo{year}{2001}), \bibinfo{pages}{319--327}.
\newblock
\urldef\tempurl%
\url{https://doi.org/10.1016/S0378-4371(01)00312-0}
\showURL{%
\tempurl}


\bibitem[\protect\citeauthoryear{Reagans}{Reagans}{2011}]%
        {reagans}
\bibfield{author}{\bibinfo{person}{Ray Reagans}.}
  \bibinfo{year}{2011}\natexlab{}.
\newblock \showarticletitle{Close encounters: Analyzing how social similarity
  and propinquity contribute to strong network connections}.
\newblock \bibinfo{journal}{\emph{Organ. Sci.}} \bibinfo{volume}{22},
  \bibinfo{number}{4} (\bibinfo{year}{2011}), \bibinfo{pages}{835--849}.
\newblock
\urldef\tempurl%
\url{https://www.jstor.org/stable/20868899}
\showURL{%
\tempurl}


\bibitem[\protect\citeauthoryear{Shirado, Fu, Fowler, and Christakis}{Shirado
  et~al\mbox{.}}{2013}]%
        {shirado2013quality}
\bibfield{author}{\bibinfo{person}{Hirokazu Shirado}, \bibinfo{person}{Feng
  Fu}, \bibinfo{person}{James~H Fowler}, {and} \bibinfo{person}{Nicholas~A
  Christakis}.} \bibinfo{year}{2013}\natexlab{}.
\newblock \showarticletitle{Quality versus quantity of social ties in
  experimental cooperative networks}.
\newblock \bibinfo{journal}{\emph{Nature communications}} \bibinfo{volume}{4},
  \bibinfo{number}{1} (\bibinfo{year}{2013}), \bibinfo{pages}{1--8}.
\newblock
\urldef\tempurl%
\url{https://doi.org/10.1038/ncomms3814}
\showURL{%
\tempurl}


\bibitem[\protect\citeauthoryear{Snijders, Van~de Bunt, and Steglich}{Snijders
  et~al\mbox{.}}{2010}]%
        {snijders2010introduction}
\bibfield{author}{\bibinfo{person}{Tom~AB Snijders}, \bibinfo{person}{Gerhard~G
  Van~de Bunt}, {and} \bibinfo{person}{Christian~EG Steglich}.}
  \bibinfo{year}{2010}\natexlab{}.
\newblock \showarticletitle{Introduction to stochastic actor-based models for
  network dynamics}.
\newblock \bibinfo{journal}{\emph{Social networks}} \bibinfo{volume}{32},
  \bibinfo{number}{1} (\bibinfo{year}{2010}), \bibinfo{pages}{44--60}.
\newblock
\urldef\tempurl%
\url{https://doi.org/10.1016/j.socnet.2009.02.004}
\showURL{%
\tempurl}


\bibitem[\protect\citeauthoryear{Sunstein}{Sunstein}{2018}]%
        {sunstein2018legal}
\bibfield{author}{\bibinfo{person}{Cass~R Sunstein}.}
  \bibinfo{year}{2018}\natexlab{}.
\newblock \bibinfo{booktitle}{\emph{Legal reasoning and political conflict}}.
\newblock \bibinfo{publisher}{Oxford University Press},
  \bibinfo{address}{Oxford, UK}. 1--42 pages.
\newblock


\bibitem[\protect\citeauthoryear{T{\'o}th, Wachs, Di~Clemente, Jakobi,
  S{\'a}gv{\'a}ri, Kert{\'e}sz, and Lengyel}{T{\'o}th et~al\mbox{.}}{2021}]%
        {toth2021inequality}
\bibfield{author}{\bibinfo{person}{Gerg{\H{o}} T{\'o}th},
  \bibinfo{person}{Johannes Wachs}, \bibinfo{person}{Riccardo Di~Clemente},
  \bibinfo{person}{{\'A}kos Jakobi}, \bibinfo{person}{Bence S{\'a}gv{\'a}ri},
  \bibinfo{person}{J{\'a}nos Kert{\'e}sz}, {and} \bibinfo{person}{Bal{\'a}zs
  Lengyel}.} \bibinfo{year}{2021}\natexlab{}.
\newblock \showarticletitle{Inequality is rising where social network
  segregation interacts with urban topology}.
\newblock \bibinfo{journal}{\emph{Nature communications}} \bibinfo{volume}{12},
  \bibinfo{number}{1} (\bibinfo{year}{2021}), \bibinfo{pages}{1--9}.
\newblock
\urldef\tempurl%
\url{https://doi.org/10.1038/s41467-021-21465-0}
\showURL{%
\tempurl}


\bibitem[\protect\citeauthoryear{Xie, Sreenivasan, Korniss, Zhang, Lim, and
  Szymanski}{Xie et~al\mbox{.}}{2011}]%
        {Xie_2011}
\bibfield{author}{\bibinfo{person}{J. Xie}, \bibinfo{person}{S. Sreenivasan},
  \bibinfo{person}{G. Korniss}, \bibinfo{person}{W. Zhang}, \bibinfo{person}{C.
  Lim}, {and} \bibinfo{person}{B.~K. Szymanski}.}
  \bibinfo{year}{2011}\natexlab{}.
\newblock \showarticletitle{Social consensus through the influence of committed
  minorities}.
\newblock \bibinfo{journal}{\emph{Physical Review E}} \bibinfo{volume}{84},
  \bibinfo{number}{1} (\bibinfo{date}{Jul} \bibinfo{year}{2011}),
  \bibinfo{pages}{011130}.
\newblock
\urldef\tempurl%
\url{https://doi.org/10.1103/physreve.84.011130}
\showURL{%
\tempurl}


\bibitem[\protect\citeauthoryear{Yuan, Alabdulkareem, et~al\mbox{.}}{Yuan
  et~al\mbox{.}}{2018}]%
        {yuan2018interpretable}
\bibfield{author}{\bibinfo{person}{Yuan Yuan}, \bibinfo{person}{Ahmad
  Alabdulkareem}, {et~al\mbox{.}}} \bibinfo{year}{2018}\natexlab{}.
\newblock \showarticletitle{An interpretable approach for social network
  formation among heterogeneous agents}.
\newblock \bibinfo{journal}{\emph{Nature communications}} \bibinfo{volume}{9},
  \bibinfo{number}{1} (\bibinfo{year}{2018}), \bibinfo{pages}{1--9}.
\newblock
\urldef\tempurl%
\url{https://doi.org/10.1038/s41467-018-07089-x}
\showURL{%
\tempurl}


\bibitem[\protect\citeauthoryear{Zhang, Lim, Korniss, and Szymanski}{Zhang
  et~al\mbox{.}}{2014}]%
        {zhang2014opinion}
\bibfield{author}{\bibinfo{person}{Weituo Zhang}, \bibinfo{person}{Chjan~C
  Lim}, \bibinfo{person}{Gyorgy Korniss}, {and} \bibinfo{person}{Boleslaw~K
  Szymanski}.} \bibinfo{year}{2014}\natexlab{}.
\newblock \showarticletitle{Opinion dynamics and influencing on random
  geometric graphs}.
\newblock \bibinfo{journal}{\emph{Scientific reports}} \bibinfo{volume}{4},
  \bibinfo{number}{1} (\bibinfo{year}{2014}), \bibinfo{pages}{1--9}.
\newblock
\urldef\tempurl%
\url{https://10.1038/srep05568}
\showURL{%
\tempurl}


\end{thebibliography}

\end{document}